\documentclass[showpacs,preprintnumbers,amsmath,amssymb]{revtex4}

\usepackage{graphicx}
\usepackage{dcolumn}
\usepackage{bm}
\usepackage{longtable}
\usepackage{rotating}
\usepackage[dvips]{color}

\begin{document}


\title{Optimization of  Generalized Multichannel Quantum Defect reference functions for Feshbach resonance characterization}

\author{R Oss\'eni}
\address{Laboratoire Aim\'{e} Cotton, CNRS, B\^at. 505, Univ Paris-Sud 11, F-91405 Orsay Cedex, France}

\author{O Dulieu}
\address{Laboratoire Aim\'{e} Cotton, CNRS, B\^at. 505, Univ Paris-Sud 11, F-91405 Orsay Cedex, France}

\author{M Raoult}
\address{Laboratoire Aim\'{e} Cotton, CNRS, B\^at. 505, Univ Paris-Sud 11, F-91405 Orsay Cedex, France}

\begin{abstract}
This work stresses the importance of the choice of the set of reference functions in the Generalized Multichannel Quantum Defect Theory to analyze the location and the width of Feshbach resonance occurring in collisional cross-sections. This is illustrated on the photoassociation of cold rubidium atom pairs, which is also modeled using the Mapped Fourier Grid Hamiltonian method combined with an optical potential. The specificity of the present example lies in a high density of quasi-bound states (closed channel) interacting with a dissociation continuum (open channel). We demonstrate that the optimization of the reference functions leads to quantum defects with a weak energy dependence across the relevant energy threshold. The main result of our paper is that the agreement between the both theoretical approaches is achieved only if optimized reference functions are used.
\end{abstract}
\maketitle

\section{Introduction}

The idea of quantum defect in atomic physics has been introduced many years ago for Rydberg atoms, to describe their similarity and difference with the hydrogen atom: the presence of the ionic core in the Rydberg atom induces a phase shift, i.e. a defect, on the wave function of the Rydberg electron which occasionally travels through the core, compared to the pure coulombic wave function which is called the {\it reference} wave function. It is well known that the usual Rydberg law for hydrogen level energies $E_{n}=-R_{\infty}/n^2$, where $R_{\infty}$ is the Rydberg constant and $n$ the principal quantum number, is then modified according to $E_{n \ell}=-R_{\infty}/(n-\mu_{\ell})^2$, where $\mu_{\ell}$ is the quantum defect associated to a Rydberg series (referred to as a {\it channel}) characterized by the orbital angular momentum $\ell$ of the outer electron. The Rydberg law can be easily extended to Rydberg electronic states of a diatomic molecule, by defining a quantum defect $\mu_{\ell,\lambda,\alpha^+}(R)$ which now depends on the projection $\lambda$ of $\ell$ on the molecular axis, on the internuclear distance $R$, and on the internal quantum state $\alpha^+$ of the molecular ionic core. In a Rydberg system, the wave function of the Rydberg electron inside the region defined by a range $r_0$ of strong interaction with the ionic core is independent, apart from a normalization factor, from the generally weak binding energy $E_{n \ell \lambda \alpha^+}(R)$ of the electron. The remaining nodal structure of the wave function outside this region is only determined by the Coulomb tail of the electron-core potential. The quantum defect is basically independent of $E_{n \ell \lambda \alpha^+}(R)$, so that it can be extrapolated through the energy threshold to describe electronic continuum states, i.e. ionization. In most systems, several Rydberg series are present in the same energy range, which most likely interact together. Such situations are described with the Multichannel Quantum Defect Theory (MQDT)\cite{seaton1983,greene1985,aymar1996}, which involves quantum defects for each of the $p$ channels and coupling parameters between them, constituting the so-called $\mathbf{Y}$-matrix with dimension $p \times p$ \cite{seaton1983}.

The quantum defect concept is very appealing in the context of ultracold gases, as it expresses the same idea than the scattering length for elastic collisions between ultracold atoms. Such a collision induces a phase shift on the wave function of the relative motion of the atoms, created by their strong interaction when they lie together within a distance $R_0$. The scattering length is an effective parameter directly related to this phase shift, and characterizes the short-range interaction between the atoms. Just like in the previous case, the wave function inside $R_0$ does not depend - apart from a normalization factor - on the relative (weak) energy of the atoms, while beyond $R_0$, it is controlled by the long-range interaction of the atoms, expressed as a multipolar expansion $\sum_{k}{C_k/R^k}$ of the potential energy. Several authors worked at designing a Generalized Multichannel Quantum Defect Theory (GMQDT) to treat interaction potentials with non-coulombic asymptotic behavior \cite{greene1979,greene1982,mies1984,mies1984a}, in order to represent the vibrational spectrum of small molecules close to their dissociation limit, or the collisions between atoms for energies just above the dissociation threshold. In contrast with the pure Coulomb field problem (defining electronic Rydberg series), there are an infinite number of possibilities for the choice of the reference functions for the molecular problem (involving vibrational series). Therefore the central issue is to find reference functions adapted to an arbitrary long-range variation of interaction potentials, in order to rigourously define quantum defects and relevant channel couplings \cite{yoo1986,jungen2000}. Recently, GMQDT has been used to study Feshbach resonances in ultracold atomic binary collision \cite{mies2000,raoult2004}. These studies have shown that GMQDT accurately recovers close-coupling (CC) cross sections results \cite{mies2000}, and brings more physical insight on the underlying molecular processes.

Using the same approach than in ref.\cite{raoult2004}, we focus our attention on the characterization of the position and the width of resonances in the photoassociation of ultracold atoms, which strongly depends on the choice of reference functions. We demonstrate that the optimization of the reference functions \cite{giusti1984a,giusti1984b,cooke1985} based on a procedure introduced in ref.\cite{lecomte1987}, leads to quantum defects with a weak energy dependence across the relevant energy threshold, just like in the standard MQDT situation. We compare our results to numerical coupled-channel calculations performed in the framework of the Fourier Grid Hamiltonian method \cite{kosloff1988} combined with an optical potential approach for the resonance characterization \cite{vibok1992}. The main result of our paper is that the agreement between the both theoretical approaches is achieved only if optimized reference functions are used in the MQDT framework.

Our paper is organized as follows. We first present in section \ref{sec:rb2} the chosen physical situation, i.e. the photoassociation of ultracold rubidium atoms into mixed electronic molecular states coupled by spin-orbit interaction. We recall next the main features of its numerical description using the Mapped Fourier Grid Hamiltonian (MFGH) method with an optical potential (section \ref{sec:mfgh}), yielding vibrational energies and predissociation line widths for high-lying vibrational levels of the coupled electronic states. In section \ref{sec:gmqdt}, we present a summary of the GMQDT treatment required to calculate the position and width of resonances. We compare in section \ref{sec:results} the results obtained by the two theoretical approaches for both the discrete and the resonant spectrum. Finally we discuss the validity of the GMQDT parameters defined in previous analysis of this system, in order to shed light on their predictive power for future experiments. Atomic units will be used for distances ($a_0$=0.052917720859~nm) and energies ($2R_{\infty}=219474.63137054$~cm$^{-1}$), except otherwise stated.


\section{The $(A, b)$ coupled state system in the rubidium dimer}
\label{sec:rb2}

The coupling mediated by spin-orbit interaction of the $A^{1}\Sigma_{u}^+ $ and $b^{3}\Pi_{u}$ electronic states (hereafter referred to as the $A$ and $b$ states, or as the ($A$, $b$) coupled system) correlated to the lowest excited $^2S+^2P$ dissociation limit in alkali dimers, represents a well-known case for the breakdown of the Born-Oppenheimer approximation which strongly perturbs their spectroscopy. This is particularly true in heavy alkali dimers like Rb$_2$ \cite{amiot1999,salami2009}, which required novel methods to deperturb the spectra recorded in high-resolution molecular spectroscopic studies \cite{lisdat2001,manaa2002}, where almost all rovibrational levels exhibit a mixed singlet/triplet character up to the dissociation limits. This pattern has also been observed in heteronuclear systems involving one heavy alkali atom like NaRb \cite{tamanis2002,docenko2007}, NaCs \cite{zaharova2009}, and RbCs \cite{bergeman2003}. In the context of the photoassociation (PA) of ultracold rubidium atoms, this strong interaction results into perturbations and predissociation resonances which have been observed for the first time in ref.\cite{cline1994}. Such perturbations have been further studied experimentally in the PA spectra of Rb$_2$ \cite{bergeman2006,jelassi2006b}, and Cs$_2$ \cite{kokoouline2002,jelassi2008b}, and modeled by various means \cite{dulieu1995,kokoouline1999,kokoouline2000,kokoouline2000a,ostrovsky2001,dion2001,kokoouline2002,jelassi2006b,jelassi2008b}. In front of this intense activity, the ($A$, $b$) coupled system in Rb$_2$ appears as a suitable test case for the present study.

In Hund's case {\it a} representation, the relevant molecular potential curves $V_A(R)$ and $V_b(R)$ of Rb$_{2}$ are associated to the $A^{1}\Sigma_{u}^+ $ and $b^{3}\Pi_{u}$ electronic states correlated to the lowest excited dissociation limits $5\ ^2S+5\ ^2P$. Due to their spin-orbit coupling, they define a subspace of $0_u^+$ symmetry in Hund's case {\it c} according to the corresponding interaction Hamiltonian $V^a_{so}(0_u^{+})$:

\begin{equation}
\textbf{V}^{(a)}_{so}(0_u^{+}) =
\begin{bmatrix}
V_b-A^{on}_{so}(R)/2 & \sqrt{3/2}A^{off}_{so}(R)\\
\sqrt{3/2}A^{off}_{so}(R) & V_A
\end{bmatrix}
\label{eq:soa}
\end{equation}

At large distances, both $R$-dependent spin-orbit coupling terms $A^{on}_{so}(R)$ and $A^{off}_{so}(R)$ reach the atomic value $2\Delta E_{so}/3$, where  $\Delta E_{so}=237.6$~cm$^{-1}$ is the fine-structure splitting of the rubidium $5p$ atomic level. In the following, we neglect the atomic hyperfine structure, as well as the rotation of the molecule, and we restrict our study to the $^{85}Rb$ isotope. Furthermore, as we focus our attention on the comparison between two approaches, we will assume $A^{on}_{so}(R) \equiv A^{off}_{so}(R)=A_{so}(R)$ in the following. Equation (\ref{eq:soa}) is rewritten in the asymptotic basis where the atomic spin-orbit interaction is diagonal (i.e. at $R \rightarrow \infty$):

\begin{equation}
\textbf{V}^{as}_{so}(0_u^{+}) =
\begin{bmatrix}
V_A/3+2V_b/3-A_{so}(R)  & \sqrt{2/3}(V_A-V_b)\\
\sqrt{2/3}(V_A-V_b)& 2V_A/3+V_b/3+A_{so}(R)/2
\end{bmatrix}
\label{eq:soas}
\end{equation}

\begin{figure}
\includegraphics[width=0.8\textwidth]{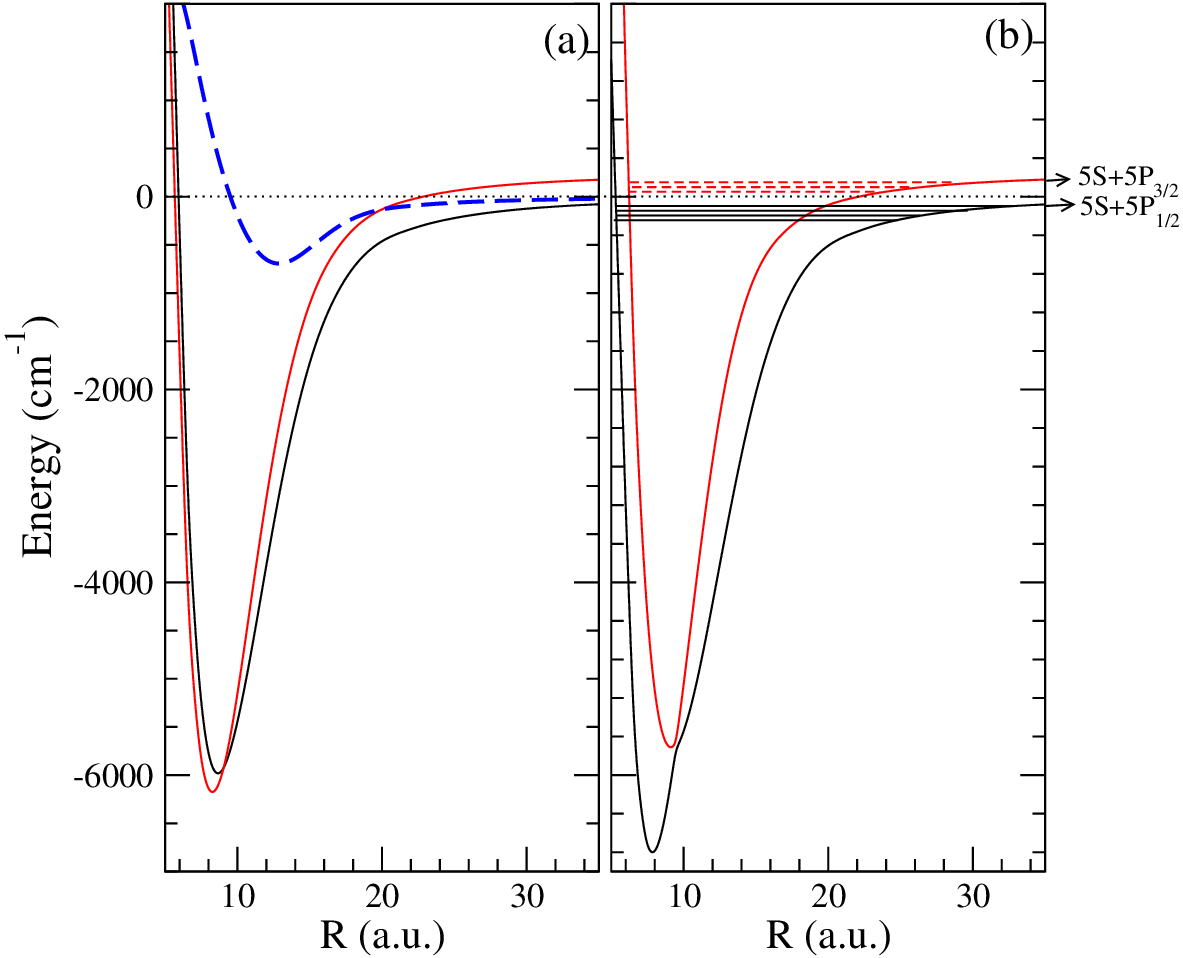}
\caption{\label{fig:potrb2} (a) Diagonal matrix elements (full lines) and coupling matrix element (dashed line) of the interaction matrix of eq.\ref{eq:soas} ({\it diabatic} representation) for the rubidium dimer; (b) Eigenvalues of the  $\textbf{V}^{as}_{so}(0_u^{+})$ matrix ({\it adiabatic} representation). For illustration purpose, bound levels and predissociating resonances are pictured with full and dashed horizontal lines, respectively. The energy origin is taken at the $5\ ^2S+5\ ^2P_{1/2}$ asymptote (horizontal dotted line).}
\end{figure}

Equation \ref{eq:soas} is convenient, as the diagonal elements now converge towards the $5^2S+5^2P_{1/2,3/2}$ dissociation limits (hereafter referred to as the $P_{1/2}$ and $P_{3/2}$ asymptotes), and the off-diagonal coupling terms vanish at finite distances. These diagonal elements and their coupling are drawn in Figure \ref{fig:potrb2}a, and represent the diabatic representation of our problem, with two potential curves encountering a real crossing around 10$a_0$, coupled by an off-diagonal term reaching its maximum around the same value. The diagonalization of $\textbf{V}^{as}_{so}(0_u^{+})$ at every $R$ yields two $0_u^+$ potential curves in the adiabatic representation (Hund's case {\it c}), with an avoided crossing around 10$a_0$ as well (Figure \ref{fig:potrb2}b). We used here the same data than in ref.\cite{kokoouline2000a}. In standard photoassociation experiments \cite{cline1994,bergeman2006,jelassi2006a}, bound levels located below one of these asymptotes are populated. Here we are interested in computing energies for truly bound levels below the $P_{1/2}$ limit, and predissociated levels lying between the $P_{1/2}$ and $P_{3/2}$ limits.

\section{The MFGH method with optical potential}
\label{sec:mfgh}

We briefly recall below the main features of the MFGH approach combined with an optical potential (MFGH+OP), which has been extensively described in several previous papers of the group \cite{kokoouline1999,kokoouline2000,kokoouline2000a,pellegrini2002,willner2004}. We want to solve the system of two coupled Schr\"odinger equations for the $(A,b)$ system expressed in the diabatic representation, which is conveniently written in a matrix form according to:

\begin{equation}
\left[\dfrac{1}{2\mu} \dfrac{d^{2}}{dR^{2}}\mathbf{I}+E \mathbf{I}-\mathbf{V}^{as}_{so}(R) \right] \mathbf{\Psi}(E,R)=0
\label{eq:psicc}
\end{equation}

where $\mathbf{I}$ is the identity matrix, $E$ the total energy, and $\mu$ the reduced mass of the rubidium dimer. In the perspective of the next section, we express the matrix $\textbf{V}^{as}_{so}$ in eq.\ref{eq:soas} with the compact form:

\begin{equation}
\textbf{V}^{as}_{so}(0_u^{+}) =
\equiv
\begin{bmatrix}
V_{oo}(R) & V_{oc}(R)\\
V_{co}(R)& V_{cc}(R)
\end{bmatrix}
\label{eq:so_oc}
\end{equation}

where the indexes $o$ hold for {\it open} and $c$ for {\it closed}, related to the $P_{1/2}$ and $P_{3/2}$ asymptotes, respectively. The solution $\mathbf{\Psi}(E,R)$ of eq. \ref{eq:psicc} is a two-component wave function $\mathbf{\Psi}(E,R) \equiv \left( \psi_o(R,E),\psi_c(R,E) \right)$ with radial wave functions $\psi_o(R,E)$ and $\psi_c(R,E)$ associated to the {\it open} and {\it closed} channels, respectively.

In the Fourier Grid Hamiltonian (FGH) approach \cite{kosloff1988,dulieu1995}, we define a grid of length $L$ with $N$ equally spaced points in the $R$ coordinate, associated with a set of $N$ plane waves $\phi_k(R)=exp(2\pi i kR/L), k=-(N/2-1),...,0,...N/2$ in the momentum space. Radial wave functions are then expressed as an expansion $\psi(R)=\sum_{i=1}^{N}{\psi(R_i)\phi_i(R)}$, in which expansion coefficients are the values of the wave functions at every grid point. The potential energy operator $\textbf{V}^{as}_{so}$ is represented by a $2N \times 2N$ matrix, composed of: (i) one $N \times N$ diagonal block per channel, each of them containing only diagonal elements equal to the potential energy of the channel at each grid point; two identical $N \times N$ off-diagonal blocks, themselves also diagonal and equal to the coupling term at each grid point. The kinetic energy operator is represented by a matrix $\textbf{T}$ which is block-diagonal, with one dense block for each of the $o$ and $c$ channels. In order to save grid points when calculating bound levels close to the dissociation limits, i.e. with large spatial extension, or predissociating levels, we define a spatial grid with a variable step size determined by the function $s(R)$ which maps the variation of the local classical kinetic energy of the radial motion in the coupled state (MFGH method \cite{kokoouline1999}):

\begin{equation}
s(R) = \frac{\pi}{\sqrt{2 \mu \left [E_{o}^{d}-V_{inf}(R)+E_{\infty} \right ]}}
\label{eq:mapping}
\end{equation}

where $V_{inf}(R)$ is the curve built from the lowest of the $V_{oo}(R)$ and $V_{cc}(R)$ potential energy values at every $R$, $E_{o}$ the energy of the lowest asymptote, i.e. the $P_{1/2}$ one, and $E_{\infty}$ the range of dissociation energies above $E_{o}$ to be explored in the numerical application.

The diagonalization of the full hamiltonian matrix $\textbf{H}=\textbf{T}+\textbf{V}^{as}_{so}$ yields $2N$ eigenvalues and eigenfunctions for the coupled system, the latter having themselves $2N$ components in this representation. In the energy range between the two dissociation limits $P_{1/2}$ and $P_{3/2}$, levels from the upper ({\it closed}) channel interact with the dissociation continuum of the lower ({\it open}) channel, inducing predissociation. Following refs.\cite{kokoouline2000a,pellegrini2002}, we add a purely imaginary potential (or optical potential) $V_{opt}(R)$ to the diagonal $V_{oo}$ term. We chose the expression proposed in ref.\cite{vibok1992}:

\begin{equation}
V_{opt}=-iA_{opt}\left[N_{opt}exp\left(-\frac{2 L_{opt}}{R-R_{opt}} \right) \right]
\label{eq:vopt}
\end{equation}

where the recommended value for the normalization factor is $N_{opt}=$13.22. The distance $R_{opt}$ characterizes the position of the optical potential, which has to be carefully chosen, depending on its amplitude $A_{opt}$ and its range $L_{opt}$. Indeed, $V_{opt}(R)$ must be placed at distances well outside the range of molecular potentials, in the region where the potential energies and the couplings are negligible. We then ensure the convergence of the accumulated phase of the predissociated levels, and of their width. Values of $L_{opt}=40a_0$ and $A_{opt}$=0.00004~a.u. were found satisfactory, while we placed the optical potential at the edge of the grid, i.e. by varying its position $R_{opt}$ between 40$a_0$ and 140$a_0$, with an upper bound for the grid ending between 80$a_0$ and 180$a_0$.

\begin{figure}
\includegraphics[width=0.6\textwidth]{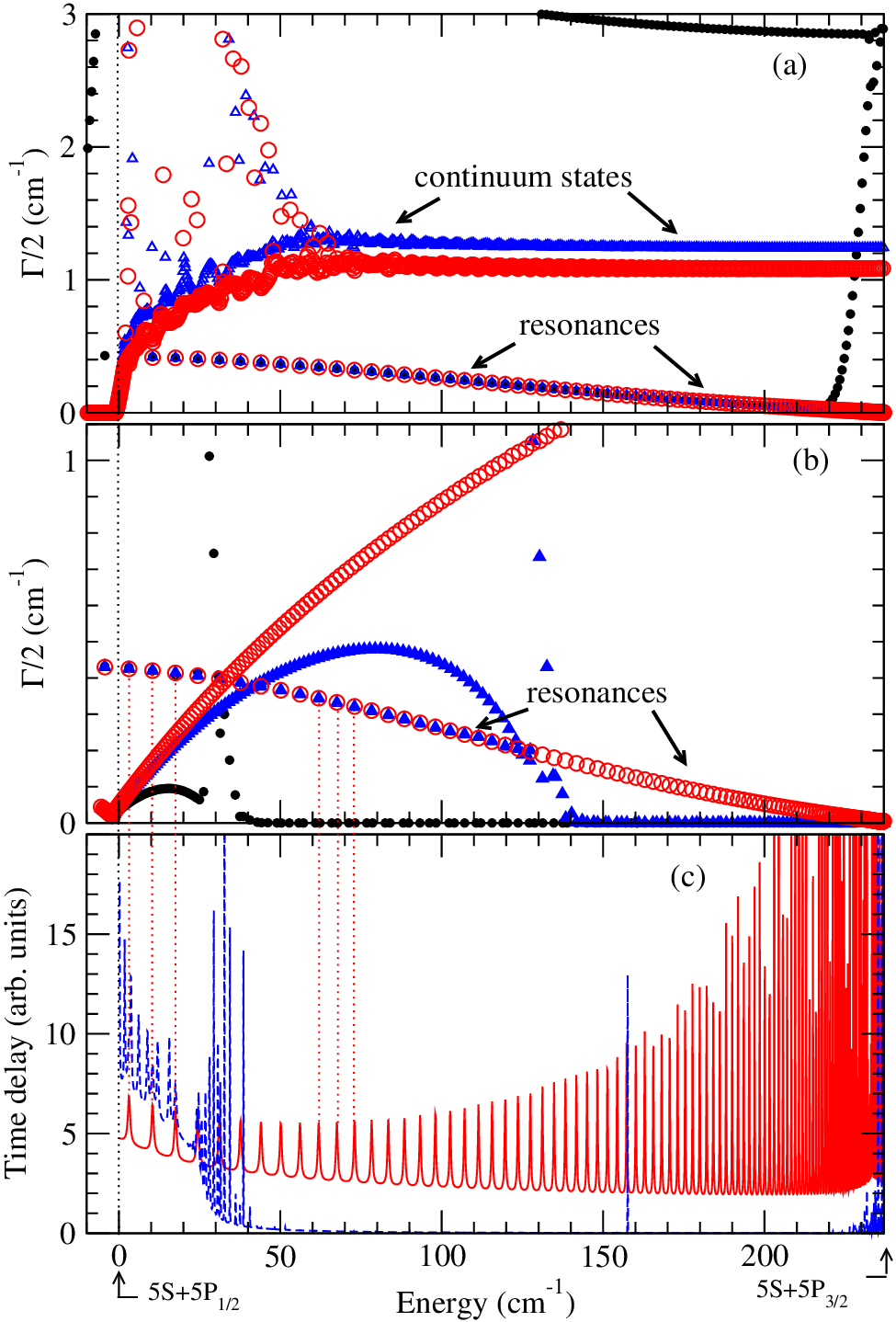}
\caption{\label{fig:convergence} (a) Imaginary part $\Gamma_k/2$ of the solutions of eq.\ref{eq:psicc} yielded by the MFGH+OP method, as a function of their real part $E_k$, between the $P_{1/2}$ and $P_{3/2}$ asymptotes. $E_{infty}$ is chosen to match these limits, and $L_{opt}=40$~a.u. and $A_{opt}$=0.00004~a.u.. Full black circles:  $L=$80 a.u. and $R_{opt}=$40~a.u.; Open blue triangles: $L=$160 a.u. and $R_{opt}=$120~a.u.; Open red circles: $L=$180 a.u. and $R_{opt}=$140 a.u.. Predissociated levels are assigned to the converged $\Gamma_k/2$ values. (b) Same as (a) with $L=$160 a.u. and $R_{opt}=$120~a.u., for different values of the $E_{\infty}=$0 (black open circles), 50~cm$^{-1}$ (blue closed triangles), 238~cm$^{-1}$ (red open diamonds). (c) Time delay {\bf which units????} associated to the complex eigenvalues, calculated according to eq.\ref{eq:delay}, for $E_{\infty}=$8~cm$^{-1}$ (blue dashed lines), and $E_{\infty}=$238~cm$^{-1}$ (red full lines). Several resonances are identified with vertical red dotted lines, for illustration purpose.}
\end{figure}

The diagonalization of the resulting complex Hamiltonian yields complex eigenvalues $\overline{E_k}=E_k-i \Gamma_k/2$, with $k=0,...,2N$. Below the $P_{1/2}$ energy, all eigenenergies correspond to bound levels with pure real energies $E_b$. Between the $P_{1/2}$ and $P_{3/2}$ asymptotes, the main task is to locate, among all complex $\overline{E}$ values, those which are associated to quasibound, or resonant levels with energy $E_r$, and a finite predissociation lifetime $1/\Gamma_r$. This can be done by several ways, illustrated in Figure \ref{fig:convergence}. As in ref.\cite{pellegrini2002}, we rely on a stabilization procedure, which consists in increasing the size of the grid, and therefore of the density of discretized states representing the continuum. The imaginary part of resonant eigenvalues $\overline{E_r}$, i.e. their width $\Gamma_r$, should converge with increasing grid size, in contrast to those associated to continuum states (see Figure \ref{fig:convergence}a). We see that resonances are identified over most of the $P_{1/2}$ and $P_{3/2}$ interval, with $L=$80 a.u. and $R_{opt}=$40~a.u.. Increasing the size of the grid indeed provide converged width over the entire energy range. We note that the use of the mapping procedure is particularly appropriate under such a circumstance where large grids are involved. In Fig. \ref{fig:convergence}b, the width of the resonances is well defined for resonances with energy lower than the chosen $E_{\infty}$ value. We see that all resonances located between the two asymptotes are well described only if $E_{\infty}$ matches the atomic spin-orbit splitting. In other words, the dissociation continuum has to be properly handled by the chosen grid step to obtain converged predissociation width. Finally, another convenient analysis of the results is provided by starting from the Breit-Wigner law for the phase shift:

\begin{equation}
\tan \delta_k = -\frac{(\Gamma_k/2)}{(E-E_{k})}
\label{eq:bwlaw}
\end{equation}

which allows defining the classical time delay \cite{smith1960a,smith1960b,fourre1994}:

\begin{equation}
\frac{\partial \delta}{\partial E} = \sum_{k} \frac{(\Gamma_k/2)}{(E-E_{k})^{2}+(\Gamma_k/2)^{2}}
\label{eq:delay}
\end{equation}

where the summation over $k$ concerns complex eigenvalues with real part larger than the energy of the $P_{1/2}$ asymptote. From Fig.\ref{fig:convergence}c, we clearly identify resonances by their larger time delay, compared to the continuum states. As expected, it is crucial to match $E_{\infty}$ to the energy interval between $P_{1/2}$ and $P_{3/2}$ to describe time delays for all resonances.

\section{GMQDT Treatment}
\label{sec:gmqdt}

We focus on a system of one {\it open} channel and one one {\it closed} channel (labeled by $o$ and $c$ indexes respectively) coupled together by a coupling vanishing beyond $R > R_{0}$. Following the previous section, they correspond to the electronic states correlated to $P_{1/2}$ ({\it open}) and $P_{3/2}$ ({\it closed}). The Schr\"odinger equation has two linearly independent solutions, which can be written beyond $R_0$ according to the following matrix form:

\begin{equation}
\mathbf{\Psi}(E,R)=\mathbf{f}(E,R) -\mathbf{g}(E,R) \mathbf{Y}(E),R > R_{0}
\label{eq:phiqdt}
\end{equation}

In this equation, all boldfaced symbols are squared matrices: $\mathbf{\Psi}(E,R)$ is the set of two independent solutions, and $\mathbf{f}$ and $\mathbf{g}$ are diagonal matrices. Their respective diagonal elements $\left(\mathbf{f}_o(R,E),\mathbf{f}_c(R,E) \right)$ and $ \left(\mathbf{g}_o(R,E),\mathbf{g}_c(R,E) \right)$ are the sets regular and irregular energy-normalized reference functions associated to the open and closed channels. The matrix elements of $\mathbf{Y}(E)$ are the phase shifts induced by the short-range interaction, parameterized by the quantum defect matrix according to $\mathbf{Y}(E)=\mathbf{tan \mu}(E)$. This equation clearly emphasizes the link between the choice of reference functions, which is not unique, and the definition of the quantum defect. For instance, when the interchannel interaction is weak one obvious choice are the solutions of the Schr\"odinger equation for the individual open and closed channel, involving the $V_{oo}$ and $V_{cc}$ diagonal matrix elements of $\textbf{V}^{as}_{so}$. In contrast, when the interchannel interactions is strong a better choice is the functions associated with the adiabatic potentials resulting from the diagonalization of $\textbf{V}^{as}_{so}$. Other possibilities have been discussed in the literature, like the analytic reference functions specific to $\frac{1}{R^{3}}$ or $\frac{1}{R^{6}}$ long-range potentials \cite{gao1998,gao2005}, or the numerical Milne solution \cite{milne1930,korsch1977} of the long-range potential as proposed by Greene and coworkers \cite{yoo1986}. The common property of all these reference functions is that the associated $\mathbf{Y}(E)$ matrix of equation (\ref{eq:phiqdt}) does not present any discontinuity across the dissociation threshold. Whatever the choice of reference functions is, the method yields the bound states energies and cross-sections, while the characterization of resonances strongly depends on this choice, as we will see below.

Let us start with the GMQDT treatment described in ref.\cite{raoult1990}, for which we outline its main steps below. Assuming a strong interaction between the open and closed channels pictured in Figure \ref{fig:potrb2}a, we determine the reference functions $\mathbf{f}^{adia}$ and $\mathbf{g}^{adia}$ associated to the individual adiabatic channels of Figure \ref{fig:potrb2}b with the Milne phase-amplitude method \cite{milne1930}. In order to avoid the tricky numerical evaluation of the non-Born-Oppenheimer coupling between the adiabatic channels, the set of coupled equations of eq.\ref{eq:psicc}) in the diabatic representation is solved by propagating the related diabatic logarithmic derivative matrix $\mathbf{L}^{dia}$ of the wave function $\mathbf{\Psi}$ using the renormalized Numerov method implemented by Johnson \cite{johnson1978}. If $\mathbf{M}(R)$ is the eigenvectors matrix of $\mathbf{H}_{so}$, it can be shown that the corresponding adiabatic logarithmic derivative matrix $\mathbf{L}^{adia}$ is given by:

\begin{equation}
\mathbf{L}^{adia}(R)=\mathbf{M}(R)~\mathbf{L}^{dia}(R)~\mathbf{M}^{t}(R)
\label{eq:logd}
\end{equation}

The coupling between the adiabatic channels is localized around their avoided crossing (see Figure \ref{fig:potrb2}b). Outside the interaction region, i.e for $R>R_{0}=13$a.u. in the present case, the $\textbf{Y}^{adia}$ matrix elements no longer depend on $R$. Therefore equation (\ref{eq:phiqdt}) holds, and $\textbf{Y}^{adia}$ is extracted according to:

\begin{equation}
\mathbf{Y}^{adia}=\left[\mathbf{L}^{adia}(R_{0})\mathbf{g}^{adia}(R_{0}) -(\mathbf{g}^{adia})^{'}(R_{0})\right]^{-1}~\left[\mathbf{L}^{adia}(R_{0}) \mathbf{f}^{adia}(R_{0})-(\mathbf{f}^{adia})^{'}(R_{0})\right]
\label{eq:yadia}
\end{equation}

The extracted $\mathbf{Y}^{adia}$ matrix involves both the closed and open channels, and relevant asymptotic conditions are applied according to the spectral range of interest. For energies below the $P_{1/2}$ asymptote, bound states energies $E_b$ are determined by solving the equation:

\begin{equation}
\vert \mathbf{tan\nu}_{c}(E_b) +\mathbf{Y}^{adia}_{cc}(E_b) \vert =0
\label{eq:det}
\end{equation}

where the vertical bars hold for the determinant of the matrix. In this expression, $\mathbf{tan\nu}_{c}$ is the {\it closed} block of the diagonal matrix associated to the accumulated phase $\nu(E)$ of the adiabatic channels, which is determined numerically using the Milne method.

In the energy range between the two asymptotes, the long-range behavior of the closed and the open channels are different. As implemented in ref.\cite{seaton1983}, the  scattering matrix $\mathbf{S}$ is obtained by applying the relevant asymptotic conditions to first extract the {\it open} block of the reaction matrix $\mathbf{K}$ defined by:

\begin{equation}
\mathbf{K}_{oo}(E)=\mathbf{Y}^{adia}_{oo}-\mathbf{Y}^{adia}_{oc} \left[\mathbf{tan\nu}_{c}(E)+ \mathbf{Y}^{adia}_{cc} \right]^{-1} \mathbf{Y}^{adia}_{co}
\label{eq:K_oo}
\end{equation}

The $\mathbf{S}$ matrix is then restricted by definition to the {\it open} channels:

\begin{equation}
{\mathbf S}_{oo}(E)=\mathbf{exp}(+i~\xi(E))_o \left[\mathbf{I}_o+i~\mathbf{K}_{oo}(E) \right]
\left[ \mathbf{I}_o-i~\mathbf{K}_{oo}(E) \right]^{-1} \mathbf{exp}(+i~\xi(E))_o
\label{eq:smat}
\end{equation}

where $(\xi(E))_o$, are the usual shift of the reference functions with respect to the Bessel reference functions, for the open channels. We have omitted the energy dependence of ${\bf Y}$ matrix on purpose, to  emphasize that the energy dependence of the scattering matrix is mainly governed by the energy variation of the parameters $\xi(E)$ and $\nu(E)$ and that the resonances  arise from the pole structure of equations (\ref{eq:K_oo},\ref{eq:smat}). The ${\bf Y}$ matrix elements are expected to slowly vary with the energy, provided that the choice of reference functions is appropriate. On the numerical side, they must be evaluated on a thin energy mesh while the ${\bf Y}$ matrix elements only require to be evaluated on a coarse mesh followed by a spline interpolation.


\section{Comparison of MFGH and GMQDT results}
\label{sec:results}

\begin{figure}
\includegraphics[width=0.6\textwidth]{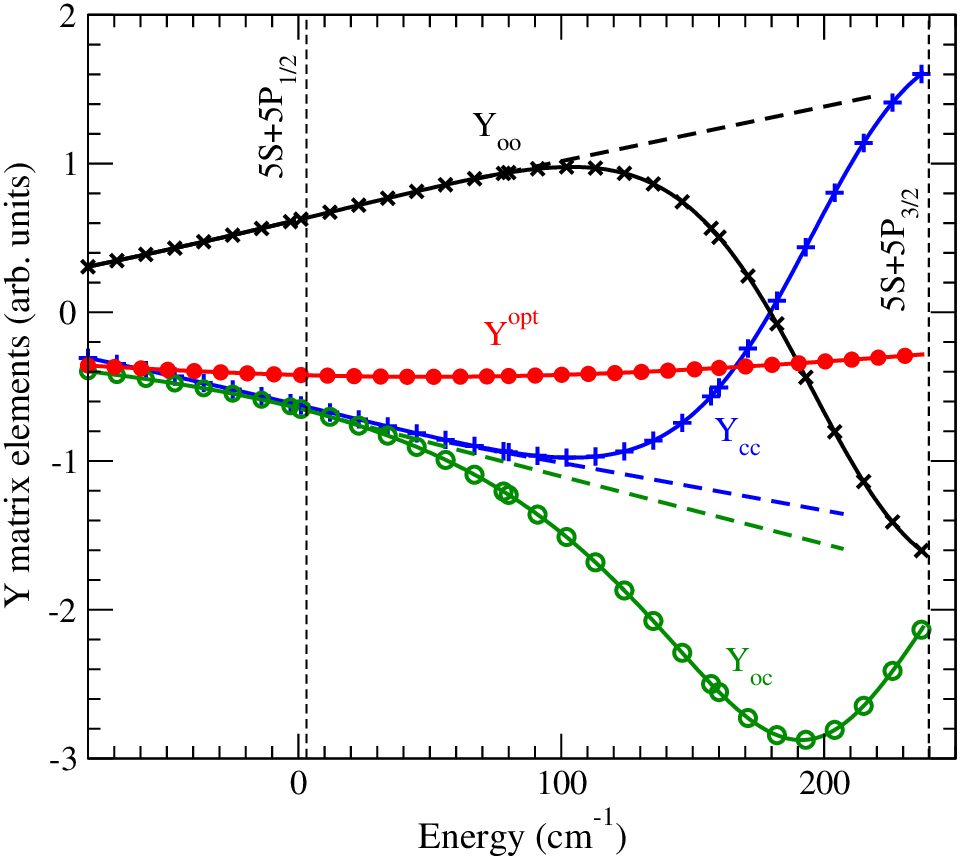}
\caption{\label{fig:ymate} Energy variation of the $Y^{adia}_{oo}$ (crosses), $Y^{adia}_{cc}$ (plus signs), and $Y^{adia}_{oc}$ (open circles) matrix elements in the region around the $P_{1/2}$ threshold. The dashed lines illustrate their linear extrapolation from energies below $P_{1/2}$ across the threshold. The off-diagonal element of the optimized matrix $Y^{opt}_{oc}$ (full circles) presents a weaker energy dependence.}
\end{figure}

In the present model, we consider the two-channel problem described in section \ref{sec:rb2}, so that there are only one open channel and one closed channel. Most of the previous GMQDT equations simplify, as the {\it open} and  {\it closed} blocks reduce to scalar quantities. First, the energy variation of the $\mathbf{Y^{adia}}$ matrix elements is displayed in Fig.\ref{fig:ymate} in the region around the $P_{1/2}$ threshold. With the present choice of reference functions associated to the adiabatic representation, it is clear that their extrapolation from the discrete range (below the threshold) through the energy continuum between the dissociation limits is rather hazardous.

\begin{figure}
\centering
\includegraphics[width=0.6\textwidth]{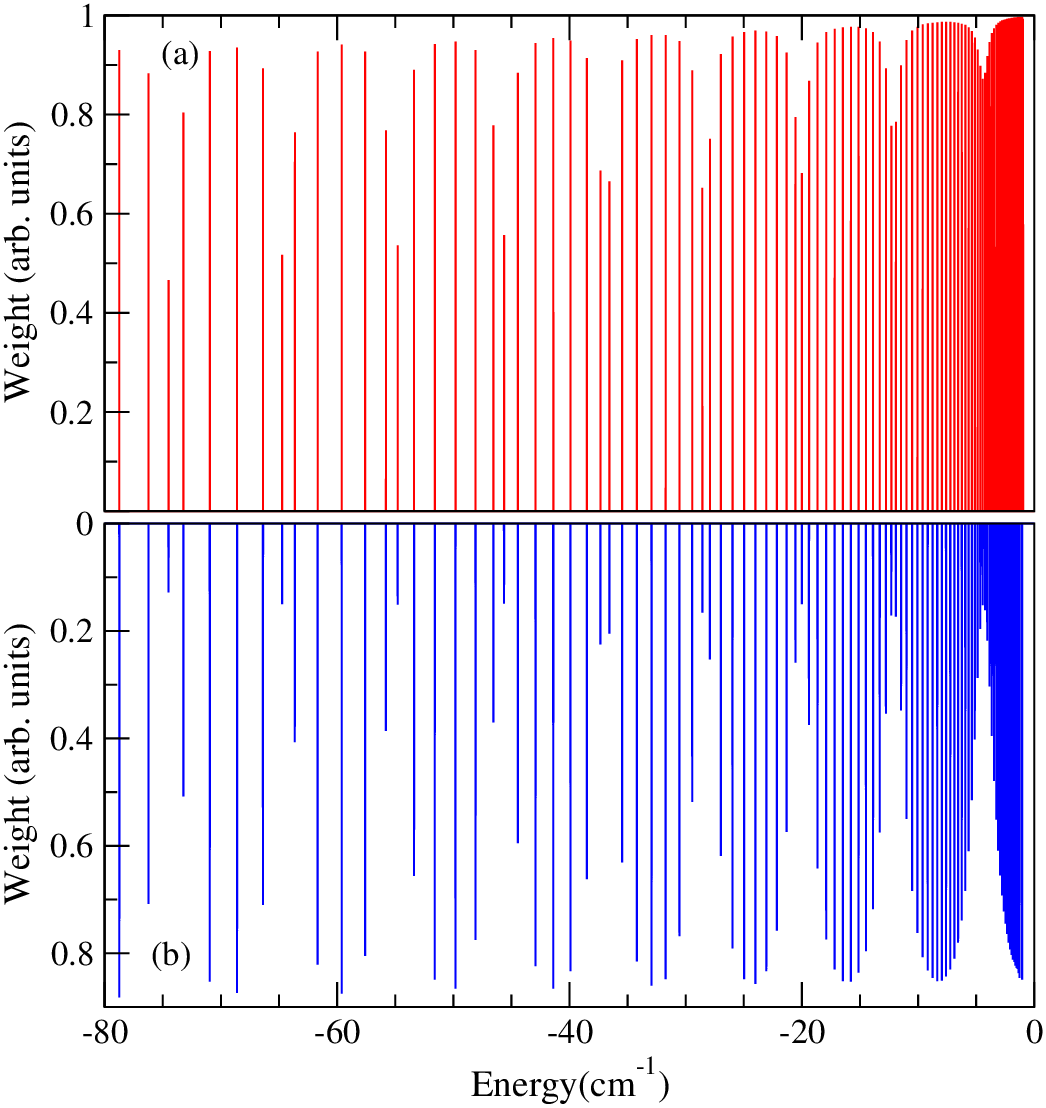}
\caption{\label{fig:mfgvsgmqdt.belowt} Weight of the lowest channel in the total wave function (see definition in the text) for bound states located below the $P_{1/2}$ limit, as a function of their energy position $E_b$: (a) by the MFGH approach and (b) by the GMQDT approach.}
\end{figure}

Before addressing the issue of the choice of the reference functions, it is useful to analyze the results obtained with the GMQDT approach in the adiabatic representation, compared to those of the MFGH method of Section \ref{sec:mfgh}. The bound state energies $E_b$ of eq.\ref{eq:det} are reported in Figure \ref{fig:mfgvsgmqdt.belowt}(b), compared to those obtained by MFGH over a 80~$cm^{-1}$ energy range below the $P_{1/2}$ threshold. In this figure, each energy value is given an ordinate corresponding to the squared modulus of the wave function for the $P_{1/2}$ channel, hereafter referred to as the {\it weight} of the lowest channel in the total two-channel wave function. The influence of the bound levels of the upper channel due to the short-range coupling is clearly visible, inducing the minima in the weights of the $P_{1/2}$ components. As expected, the agreement on bound state energies is perfect between the two approaches. In contrast, it is not surprising that the weights are not identical in the two representations, as the chosen channels are not defined in the same manner, as it can be seen by comparing the two panels of Figure \ref{fig:potrb2}.

\begin{figure}
\centering
\includegraphics[width=0.6\textwidth]{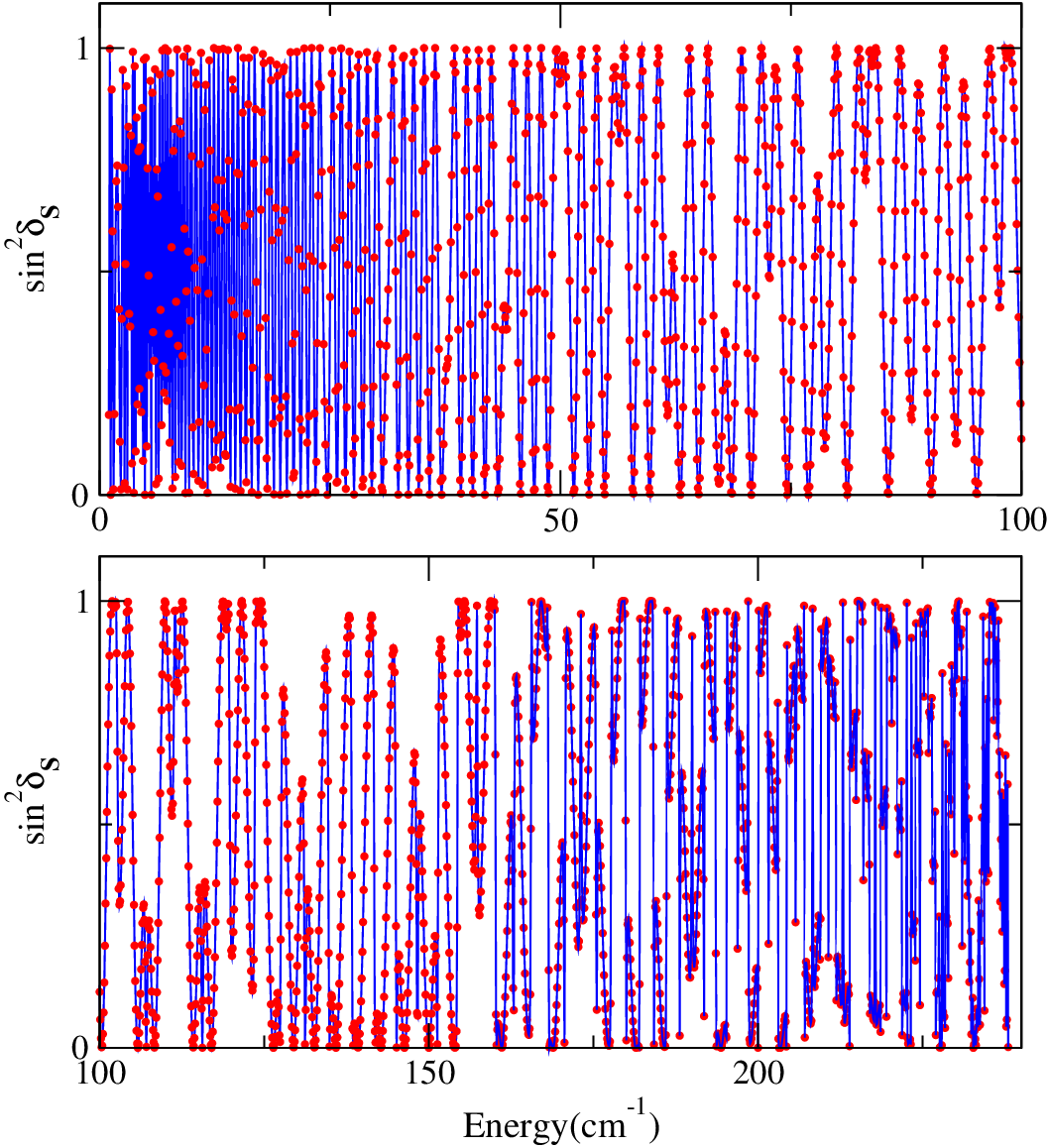}
\caption{\label{fig:smat2} Energy variation of the total phase shift $\delta_{S}$ between the $P_{1/2}$ limit (taken as the origin of energies) and the $P_{3/2}$ limit. Full lines correspond to the GMQDT calculation performed with the adiabatic set of references functions. Full circles correspond to the results obtained with the rotated set of reference functions. }
\end{figure}

Next, the results in the  energy range between the two dissociation thresholds where the resonances are expected to take place are interpreted through the $\mathbf{S}$ matrix elements, of more precisely through its eigenphases $\delta_S$ defined as $\mathbf{S} \equiv \mathbf{exp}(2i~\delta_S)$. We reported in Figure \ref{fig:smat2} the quantity $\sin^2\delta_S$ (see the full lines in the figure), which shows that its energy variation is so irregular that it is impossible to recognize any resonant pattern. This suggests that the dynamics of both channels is strongly coupled to each other. One can isolate the contribution of the resonances to the total phase shift $\delta_S$, by first setting the notation $\tan \delta_K = \mathbf{K}_{oo}$ so that $\delta_S$ can be extracted from eq. \ref{eq:smat}: $\delta_{S}=\xi_{o}+\delta_K$.

\begin{figure}
\centering
\includegraphics[width=0.6\textwidth]{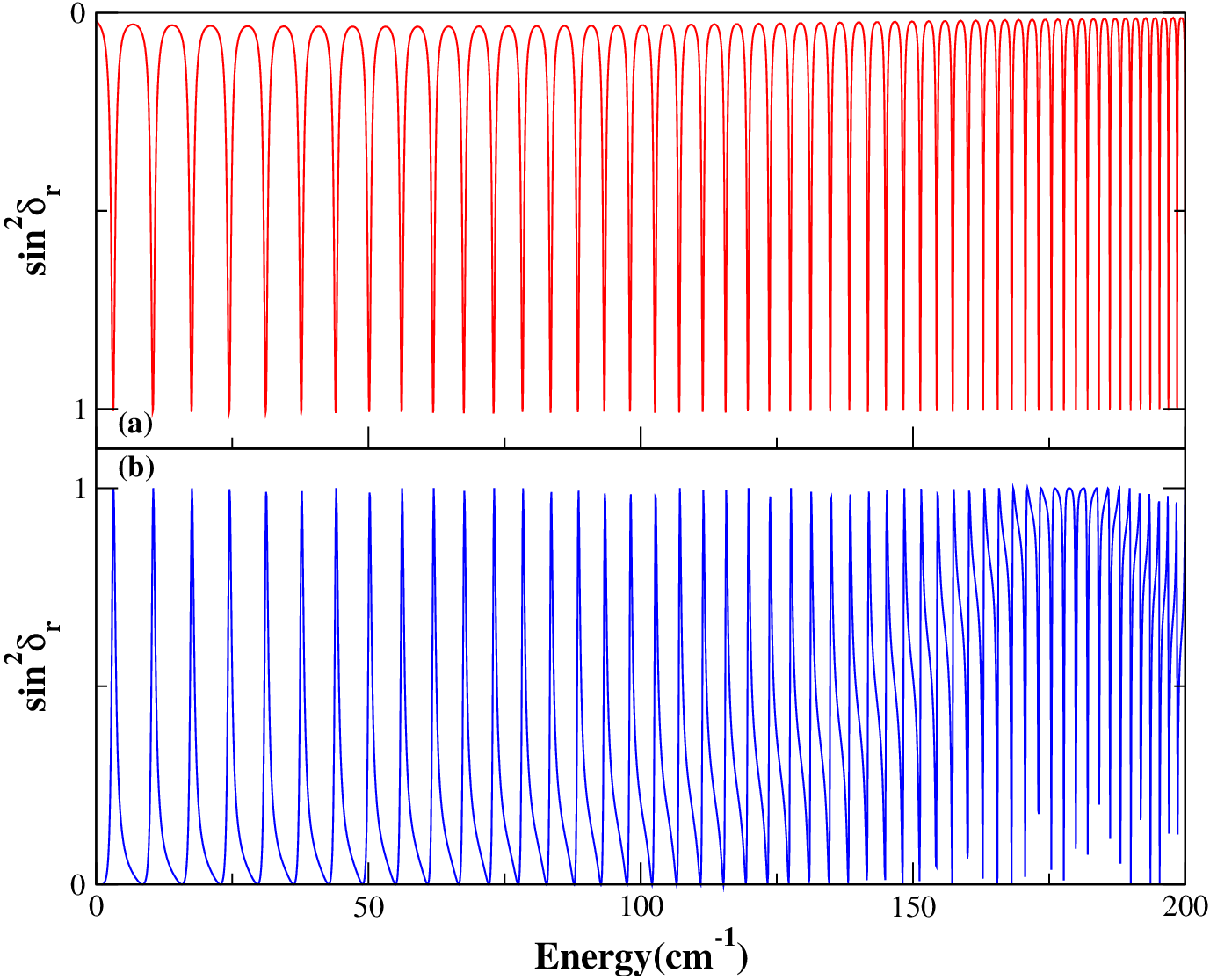}
\caption{\label{fig:sigmares.old} Energy variation of the resonant part of the phase shift $\delta_{r}$ between the $P_{1/2}$ limit (taken as the origin of energies) and the $P_{3/2}$ limit. (a) MFGH calculations; (b) GMQDT calculations.}
\end{figure}

If we neglect the channel interaction, the $\mathbf{K}_{oo}$ reduces to $\mathbf{Y}^{adia}_{oo}$ and we can define a so-called {\it background} phase shift according to $\tan \delta_{bg} = \mathbf{Y}^{adia}_{oo}$. The resonant part $\delta_{r}$ of the phase shift can be calculated as $\delta_{r}=\delta_K-\delta_{bg}$. The energy variation of $\sin^{2}\delta_{r}$ between the two asymptotes is displayed in Fig.\ref{fig:sigmares.old}a, where the regular series of resonances can now be identified from the energy of the $P_{1/2}$ limit until about 100~cm$^{-1}$ upwards. One can note that even in the low energy part the resonance profiles do not correspond to a Lorentzian shape, as we would expect from an isolated resonant contribution. A gradual increase of the profile asymmetry with increasing energy is visible, decreasing again close to the $P_{3/2}$ limit. For comparison, one can also extract this resonant contribution from the MFGH+OP approach (Fig.\ref{fig:sigmares.old}a). Resonant energies $E_r$ and width $\Gamma_r$ are deduced as outlined in Section \ref{sec:mfgh}, which assumes that the related resonances are described by a Breit-Wigner profile (eq.\ref{eq:bwlaw}) which can be expressed in an equivalent way as:

\begin{equation}
\sin^{2} \delta_{r}(E)=\dfrac{(\Gamma_{r}/2)^{2}}{(E-E_{r})^{2}+(\Gamma_{r}/2)^{2}}
\label{eq:sinbreitwigner}
\end{equation}

We see that in Figure \ref{fig:sigmares.old} both approaches yield identical resonance positions up to about  160~cm$^{-1}$ above $P_{1/2}$. However, due to the asymmetric resonance profile, the comparison of width is not obvious. Actually, the resonance width can be directly extracted from the $\mathbf{Y}^{adia}$ matrix, using a procedure proposed by Lecomte \cite{lecomte1987}. The starting point is to eliminate the open channels to build the block of the $\mathbf{K}$ matrix associated to the closed channel by applying Siegert aymptotic conditions, instead of doing the reverse as in eq.\ref{eq:K_oo} (we omitted the {\it adia} index for simplicity):

\begin{equation}
\mathbf{K}_{cc}=\mathbf{Y}_{cc}-\mathbf{Y}_{co}\mathbf{Y}_{oo}\left[\mathbf{I}+\mathbf{Y}_{oo}^{2}\right] ^{-1}\mathbf{Y}_{co}+\imath \mathbf{Y}_{co}\left[\mathbf{I}+\mathbf{Y}_{oo}^{2}\right] ^{-1}\mathbf{Y}_{oc}
\label{eq:K_cc}
\end{equation}

Following the framework of configuration interaction as described by Fano \cite{fano1961}, one can interpret all terms of this expression. The imaginary part is due to the interaction between the closed and open channel $\mathbf{Y}_{co}$ which give rises to resonance in the collisional cross-sections, and is proportional to the width of the resonance. The real part consists in two terms. The leading term is the  direct coupling term $\mathbf{Y}_{cc}$ between the closed channels. It is modified by the next term which describes the indirect coupling among the closed channels through their interaction with the open ones. This latter term represents the shift of the resonance relative to its zeroth-order position determined by the direct interaction term $\mathbf{Y}_{cc}$. The resonance energies $E_{r}$ are located by solving the determinental equation:

\begin{equation}
\tan \nu_{c}(E_{r}) +Y_{cc}-Y_{co}Y_{oo}\left[1+Y_{oo}^{2}\right] ^{-1}Y_{oc}=0
\label{eq:detkcc}
\end{equation}

which reduces to a scalar equation in our case with one single closed channel and one single open channel. The zeroth-order energy position $E^{(0)}_r$ of the resonance is obtained by canceling all interaction terms in eq.\ref{eq:detkcc} such that:

\begin{equation}
\tan \nu_c(E^{(0)}_r)+Y_{cc}=0
\label{eq:E0}
\end{equation}

Using a first-order development of $\tan \nu_{c}(E_{r}$ around $E^{(0)}_r$ in eq.\ref{eq:detkcc}, the resonant energy can easily be extracted as $E_{r}-E_{0}- \Delta=0$, with the energy shift:

\begin{equation}
\Delta = N^{-2}(E_{0})Y_{co}\left[ (Y_{oo}(I+Y_{oo}^{2})^{-1}\right] Y_{oc}
\label{eq:shift}
\end{equation}

where $N^2(E)$ is the usual QDT bound state normalization factor \cite{seaton1983}:

\begin{equation}
N^2(E)=\dfrac{\partial \nu}{\partial E}\dfrac{1}{cos^{2}\nu(E)}
\label{eq:qdtnorm}
\end{equation}

and is the usual QDT bound state normalization factor (Seaton 83). The associated width is then given by:

\begin{equation}
\Gamma_r = 2N^{-2}(E_{r}) ~Y_{co}\left[ (I+Y_{oo}^{2})\right]^{-1} Y_{oc}
\label{eq:gamaqdt}
\end{equation}

\begin{figure}
\centering
\includegraphics[width=0.6\textwidth]{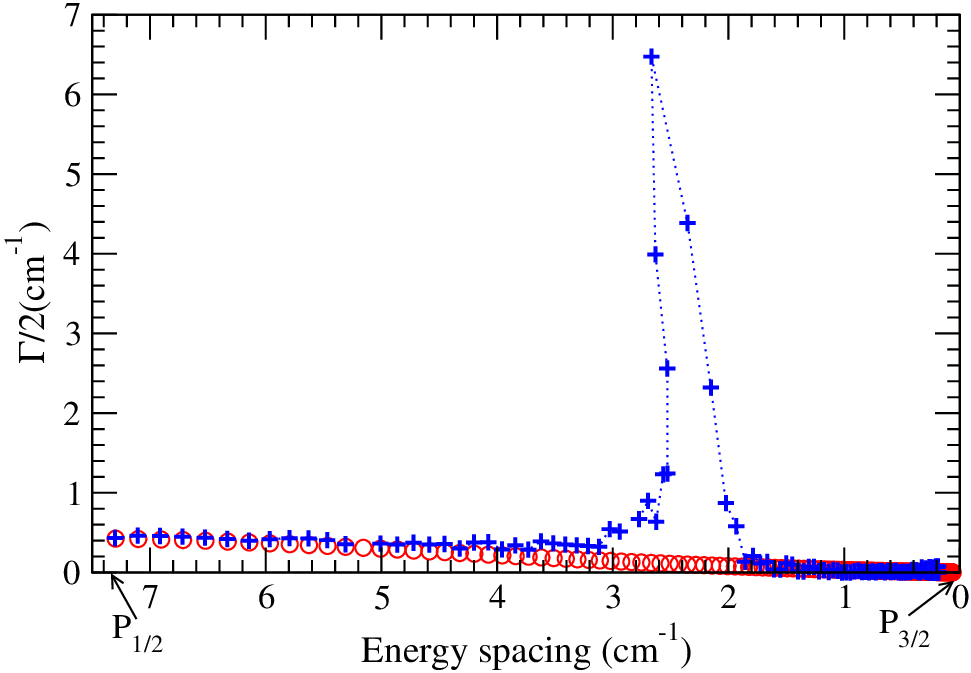}
\caption{\label{fig:gamaold} Energy variation of the resonance width given by Eq.\ref{eq:gamaqdt} as a function of the energy spacing of the resonant levels. Circles: MFGH; Plus signs: GMQDT.}
\end{figure}

The variation of the resonance widths $\Gamma_r$ with their energy $E_r$ is presented in Figure \ref{fig:gamaold}, as computed by both GMQDT and MFGH approaches. One can observe a good agreement between the two methods in the low-energy domain, while big discrepancies occur around 175~cm$^{-1}$ above $P_{1/2}$. In this region, the resonance can no longer be considered as being isolated as their width become larger than the energy separation between consecutive resonances. Therefore the expression of eq.\ref{eq:gamaqdt} no longer hold, and the apparent narrow widths computed with GMQDT no longer reflect the actual strong coupling between the two channels \cite{mies1984a}. For instance, their apparent widths in fig.\ref{fig:sigmares.old} are much smaller than the values given by Eq.\ref{eq:gamaqdt}, which usually referred to as stabilization effect. The asymmetric (non-Lorentzian) profiles result from interferences between neighboring overlapping resonances.

\section{Optimization of GMQDT reference functions}
\label{sec:optim}

The apparent failure of GMQDT to properly characterize resonances in the dissociation continuum actually arises from an improper choice of the set of reference functions. Indeed, we have seen in Fig.\ref{fig:ymate} that the intuitive choice of reference functions associated to the adiabatic channels induce large $\mathbf{Y}$ matrix elements with strong energy variation across the dissociation threshold. In other words, the non-adiabatic couplings are so strong that the adiabatic channels do not represent a good zeroth-order approximation in the present case. Several authors have shown that an appropriate rotation of the reference functions can improve such a situation. The basic idea is to define a new $\mathbf{Y}$ matrix where only the off-diagonal block is not zero. In such a representation no energy shift would occur on the resonance, i.e. the coupling between the open and closed channel would be minimized. Following ref.\cite{lecomte1987}, this goal is achieved in two steps. Two successive rotations of the initial adiabatic $f_c$ and $g_c$ (resp. $f_o$ and $g_o$) reference functions with a characteristic angle $\theta_c$ (resp $\theta_o$)are applied according to:

\begin{eqnarray}
\overline{f}_{c}&=&f_{c} \cos \theta_{c} - g_{c} \sin \theta_{c} \\
\overline{g}_{c}&=&f_{c} \sin \theta_{c} + g_{c} \cos \theta_{c}
\label{eq:fgrot}
\end{eqnarray}

with a similar equation for $f_o$ and $g_o$ with the angle $\theta_o$. After each rotation, the new matrix $\mathbf{Y}^{rot}$ is given by:

\begin{eqnarray}
\left( \begin{array}{cc}
\mathbf{Y}^{rot}_{oo}&\mathbf{Y}^{rot}_{oc} \\
\mathbf{Y}^{rot}_{co}&\mathbf{Y}^{rot}_{cc}
\end{array}
\right)
=
\left( \begin{array}{cc}
\mathbf{Y}_{oo}&\mathbf{Y}_{oc} \\
\mathbf{Y}_{co}&\mathbf{Y}_{cc}
\end{array}
\right)
\left[
\left( \begin{array}{cc}
\cos \theta_o&0 \\
0&\cos \theta_c
\end{array}
\right)
-
\left( \begin{array}{cc}
\sin \theta_o&0 \\
0&\sin \theta_c
\end{array}
\right)
\right]\\
\times
\left[
\left( \begin{array}{cc}
\mathbf{Y}_{oo}&\mathbf{Y}_{oc} \\
\mathbf{Y}_{co}&\mathbf{Y}_{cc}
\end{array}
\right)
\left( \begin{array}{cc}
\sin \theta_o&0 \\
0&\sin \theta_c
\end{array}
\right)
+
\left( \begin{array}{cc}
\cos \theta_o&0 \\
0&\cos \theta_c
\end{array}
\right)
\right]^{-1}
\label{eq:yrot}
\end{eqnarray}

with $\theta_o=0$ for the first rotation, and $\theta_c=0$ for the second one. First, the energy-dependent angle $\theta_{c}$ is chosen such that the real part of the new $\mathbf{K}^{rot}_{cc}$ matrix associated to $\mathbf{Y}^{rot}$ vanishes, leading to:

\begin{equation}
\tan 2\theta_{c}= \dfrac{2 Re(\mathbf{K}_{cc})} {1-Im(\mathbf{K}_{cc})^{2}-Re(\mathbf{K}_{cc})^{2}}
\label{eq:tetac}
\end{equation}

In this rotated reference set, eq.\ref{eq:detkcc} used to locate the resonance simply reduces to:

\begin{equation}
\tan(\nu_{c}(E)+\theta_{c})=0
\label{eq:detkccrot}
\end{equation}

For the second step,  $\mathbf{Y}^{opt}_{oo}$ is set to zero, which is fulfilled for the value $\theta_{o}$ defined by:

\begin{equation}
\tan \theta_{o}=\mathbf{Y}^{rot}_{oo}
\label{eq:tetao}
\end{equation}

It can be easily derived from Eq.\ref{eq:yrot} that this also ensures that   $\mathbf{Y}^{opt}_{cc}$ is zero. Hence, the $\mathbf{Y}^{opt}$ matrix is well transformed into a matrix with non-vanishing off-diagonal blocks.

The energy variation of this single matrix element  $\mathbf{Y}^{opt}_{oc}$ is reported in Fig.\ref{fig:ymate}. One clearly observes that the magnitude of the matrix element, associated with the rotated reference functions, is now smaller than unity. Moreover its variation with the energy is much weaker than the variation observed for the off-diagonal element associated with the non-rotated reference functions.

\begin{figure}
\centering
\includegraphics[width=0.6\textwidth]{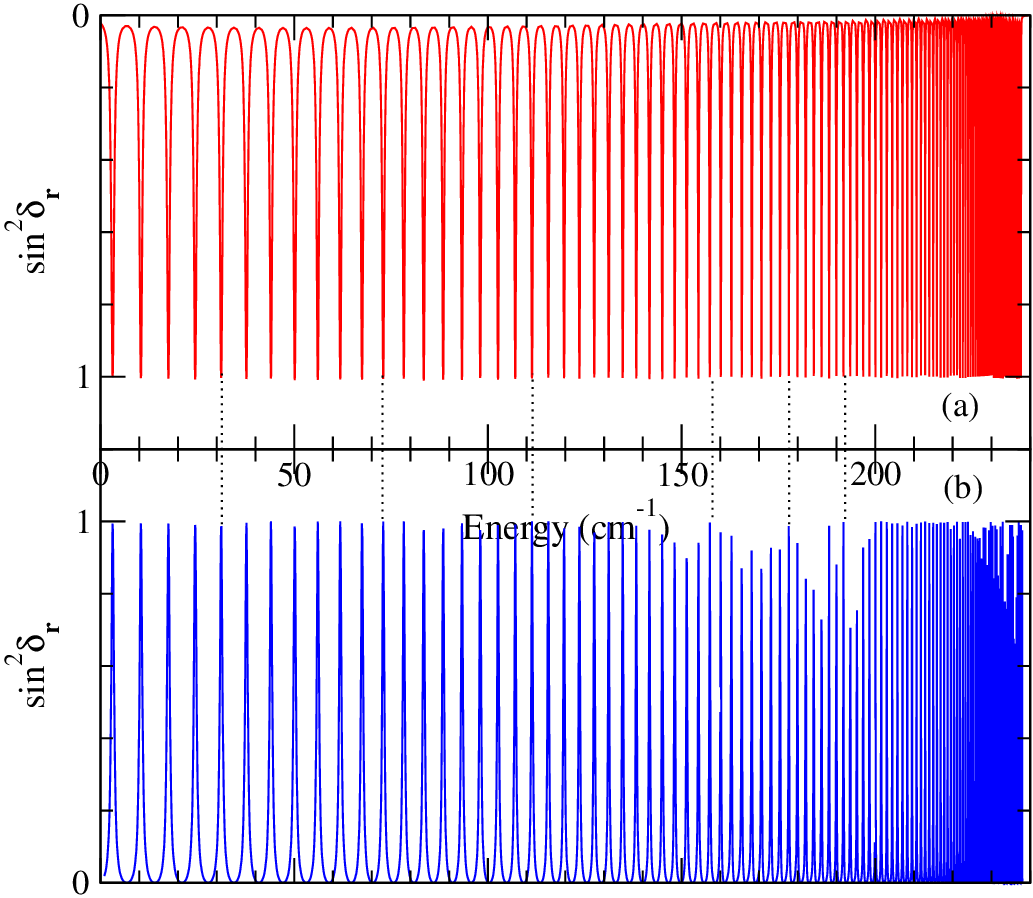}
\caption{\label{fig:sigmares} Energy variation of the resonant part of the phase shift $\delta_{r}$ between the $P_{1/2}$ limit (taken as the origin of energies) and the $P_{3/2}$ limit. (a) MFGH calculations; (b) GMQDT calculation with optimized reference functions.}
\end{figure}

The  energy variation of eigenphases of the scattering matrix determined with the optimized reference set is presented in Fig.\ref{fig:smat2} above. It must be noted that this result is obtained with eqs.\ref{eq:smat} in which $\xi_{o}$ must be replaced by $\xi_{o}+\theta_{o}$, $\nu_{c}$ by $\nu_{c}+\theta_{c}$ and the $\mathbf{Y}$ matrix by its optimized counterpart. At first glance, as it should be, this variation exactly matches the one obtained with the non-rotated reference set. Within this rotated frame, the analysis of the resonant contribution $\delta_{r}$ to the scattering matrix eigenphase $\delta_{S}$ reported in Figure \ref{fig:sigmares}, shows that the resonances now appear with the expected Lorentzian profile. Moreover both theoretical approaches leads to the same results for the resonance location and width.

\begin{figure}[t]
\centering
\includegraphics[width=0.6\textwidth]{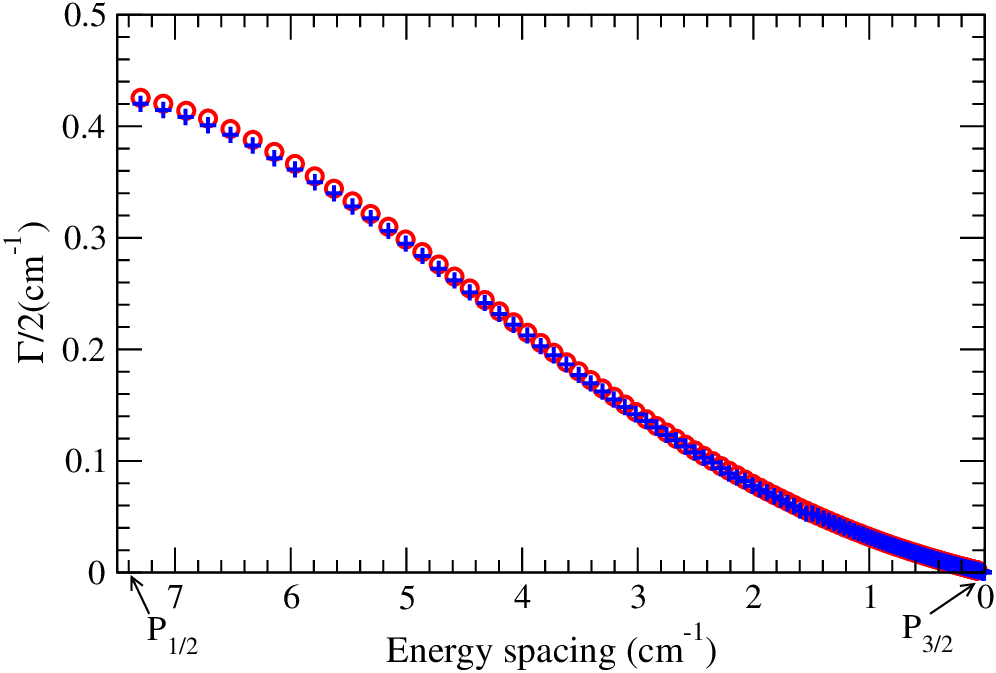}
\caption{Energy variation of the resonance width obtained from MFGH+OP (open circles), and from GMQDT using rotated reference functions (plus signs), as functions of the energy spacing between the resonant levels.}
\label{fig:gamanew}
\end{figure}

A more quantitative comparison of the width of the resonances is reported in Fig.\ref{fig:gamanew}. The agreement between GMQDT and MFGH shows that the Fermi golden rule is recovered in the rotated set of reference functions. Although the collision cross-section energy variation is independent of the choice of reference functions, this application  demonstrates how this choice is important for the analysis of resonances arising in the cross-section..

\section{Conclusion}

In this paper we emphasized on the importance of the choice of reference functions in the GMQDT treatment of collisional resonances, illustrated by the case of the photoassociation of cold atoms. In contrast with the case of interacting Rydberg series, where the  quantum defects are always associated with the regular and irregular Coulomb functions, the present two-channel molecular problem allows several choices for the reference functions. We demonstrated that the widths of the resonances can be properly assigned provided that the reference functions are chosen in a way which minimizes the interaction with the related channels, and which ensures a weak variation of the associated quantum defect with the energy. Therefore the width of the resonances can be safely deduced after the extrapolation through the dissociation threshold of the generalized quantum defects computed for the bound state spectrum. We want to stress that both set of reference functions lead to the same collisional cross-sections because they correspond to a different balance  between the background (governed by the reference functions) and the resonant contributions to the $\textbf{S}$ matrix eigenphase. They only lead to a different analysis of the resonance showing up in the cross-sections. The non-rotated set corresponds to a strong coupling situation for which it is  difficult to have predictable physical insight, while the rotated set corresponds to a more familiar and predictable weak coupling situation.

\begin{figure}[t]
\centering
\includegraphics[width=0.6\textwidth]{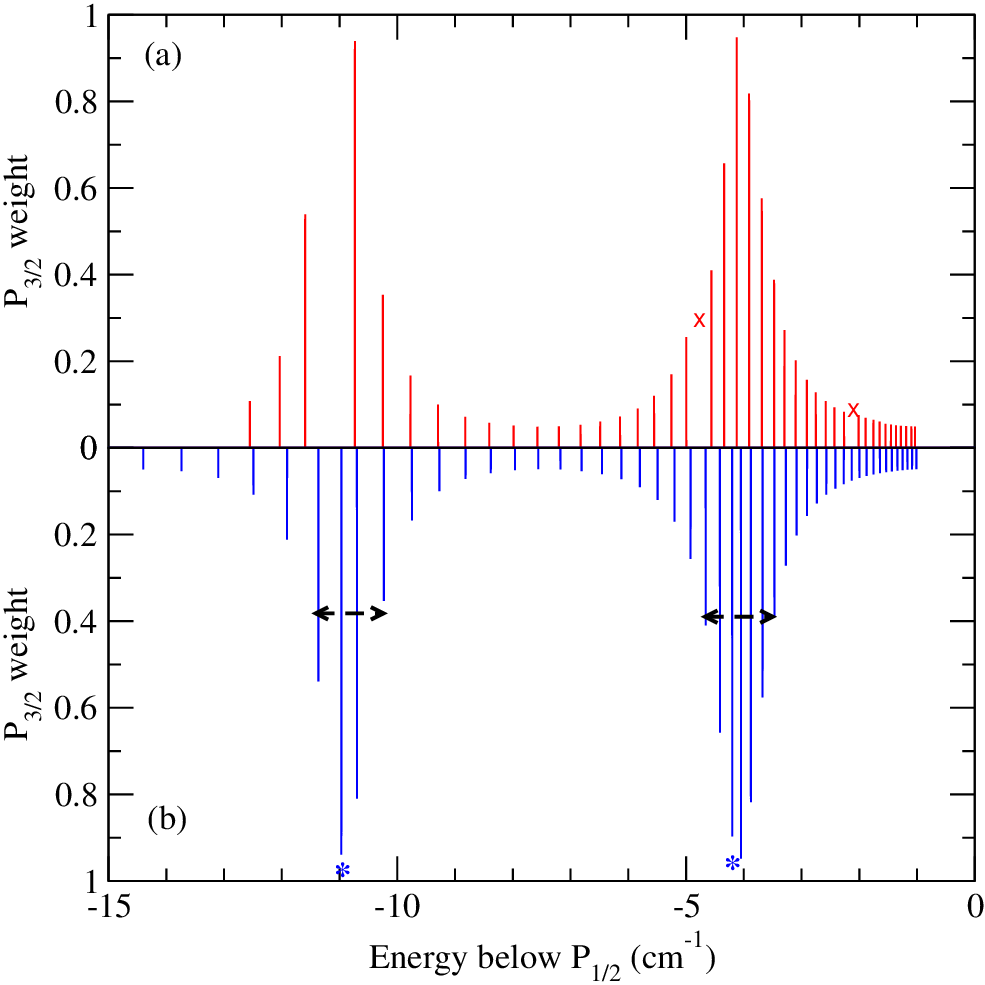}
\caption{\label{fig:rb87exp} Energies of the bound states of the $^{87}$Rb$_2$ $0_u^+ (A, b)$ coupled states computed within the GMQDT framework with optimized reference functions (see text), compared to experimental results of ref.\cite{jelassi2006b}, below the $P_{1/2}$ asymptote taken as the origin of energies. Intensities of the vertical lines represent the weight of the $P_{3/2}$ channel in the GMQDT results. Experimental results are weighted by the same quantity, for convenience. Crosses indicate missing data in the experimental signal due to technical reasons. Stars indicate levels predicted by our model, which are not detected in the experiment. Double-sided arrows indicate the approximate energy width of the perturbation induced by levels of the $P_{3/2}$ channel in the $P_{1/2}$ channel.}
\end{figure}

With this in mind, it is worthwhile to revisit the interpretation of the experimental results reported in ref.\cite{jelassi2006b} concerning the photoassociation of cold $^{87}$Rb atoms into the $0_u+ (A, b)$ coupled states. In Figure \ref{fig:rb87exp}, we reported the positions of the recorded PA lines below the $P_{1/2}$ asymptote, and the energies of the corresponding levels computed with the GMQDT approach using the rotated reference functions above. We see that the molecular potentials used for the calculations provide energies in reasonable agreement with the experimental values, even if there is room for their improvement. More specifically, within the Lu-Fano approach \cite{lu1970} used in refs.\cite{kokoouline2002,jelassi2006b}, the underlying assumption is that the reference functions are associated to the long-range adiabatic $0_u^+$ potentials varying as $R^{-3}$, even if these functions are not explicitly derived. This agreement simply means that our potential curves exhibit a long-range behavior which is close to the actual one. In this figure, the computed weight of the wave functions on the $P_{3/2}$ channel is used to yield an intensity to the lines. The resonance induced by the upper channel corresponds to the highest value of this weight. Over the displayed energy range, two of such resonances appear, separated by about 7~cm$^{-1}$. We note however that for each of these resonances, we predict the existence of one line undetected in the experiment, marked with a star in Figure \ref{fig:rb87exp}. They correspond to pairs of one $P_{3/2}$ and one $P_{1/2}$ level, which are the more strongly coupled to each other. It seems that only one partner of such pairs has been identified in the experiment.

Such a representation is helpful to define the energy range of the $P_{1/2}$ bound spectrum which is perturbed by the interloper belonging to the $P_{3/2}$ channel. Indeed, the double-sided arrows in the figure suggest that the half-width of this perturbation is equal to about 0.5~cm$^{-1}$, in agreement with the computed half-width of the very actual resonances lying just above the $P_{1/2}$ threshold reported in Figure \ref{fig:gamanew}. This value is also in agreement with the energy width of the variation of the quantum defect around the resonances given in ref.\cite{jelassi2006b}. In this latter paper, the authors extrapolate their model through the dissociation limit, which is indeed justified by the present analysis. For the first resonance above $P_{1/2}$, their quantum defect indeed varies over a half-width of about 0.4~cm$^{-1}$. However at the predicted energy position in their signal, they assigned a broader (2~cm$^{-1}$) feature to this resonance. Therefore we suggest that this feature could be revisited experimentally, as well as higher resonances, hopefully to rule out any unwanted contribution to its broadening.

Finally, predissociation resonances over about 35~cm$^{-1}$ below the $P_{3/2}$ threshold of $^{85}$Rb$_2$ have been observed in ref.\cite{bergeman2006}. Their spectrum show a series of resonances with decreasing half-widths, which can be estimated from their figure as varying from about 0.8~cm$^{-1}$ down to about 0.02~cm$^{-1}$ close to $P_{3/2}$. This trend is in agreement with the one visible in Figure \ref{fig:gamanew}, where we predict a quasi-linear decrease of the half-width in this region, from 0.06~cm$^{-1}$ down to almost zero. A more accurate determination of the half-width from raw data of ref.\cite{bergeman2006} is desirable to better compare its amplitude and variation with our prediction. It would represent a test of the spin-orbit coupling used in the calculation. In this respect, a new perspective for the present analysis is offered by the recent determination of the potential curves and spin-orbit coupling of the $(A,b)$ coupled system in Rb$_2$, extracted from the compilation of several high-resolution spectroscopic data \cite{salami2009}. Indeed, the present calculations could be run again with these new accurate data, in order to predict the predissociation width over the whole energy range  between $P_{1/2}$  and $P_{3/2}$, and hopefully confirm the interpretation of the PA results of ref.\cite{bergeman2006}.

\section*{Acknowledgments}
Fruitful discussions with Laurence Pruvost about her experimental results, and with Jean-Marie Lecomte, are gratefully acknowledged.


\end{document}